\documentclass[aps,prl,reprint,amsmath,amssymb,showpacs,superscriptaddress]{revtex4-1}
\usepackage{CJK}
\usepackage{graphicx}
\usepackage{dcolumn}
\usepackage{bm}
\usepackage{mathrsfs}   
\usepackage{color,xcolor}  
\usepackage{multirow}  
\usepackage{booktabs} 
\usepackage{mathtools}  

\usepackage{hyperref}

\newcommand{\bbm}{\begin{bmatrix}}
\newcommand{\ebm}{\end{bmatrix}}
\newcommand{\bBm}{\begin{Bmatrix}}
\newcommand{\eBm}{\end{Bmatrix}}
\newcommand{\bpm}{\begin{pmatrix}}
\newcommand{\epm}{\end{pmatrix}}

\begin{document}


\title{Tensor-force effects on nuclear matter in relativistic \textit{ab initio} theory}

\author{Sibo Wang}
\affiliation{Department of Physics and Chongqing Key Laboratory for Strongly Coupled Physics, Chongqing University, Chongqing 401331, China}

\author{Hui Tong}
\affiliation{Helmholtz institute for radiation and nuclear physics, University of Bonn, Bonn D-53115, Germany}

\author{Chencan Wang}
\affiliation{School of Physics and astronomy, Sun Yat-Sen University, Zhuhai 519082, China}

\author{Qiang Zhao}
\affiliation{Center for Exotic Nuclear Studies, Institute for Basic Science, Daejeon 34126, Korea}

\author{Peter Ring}
\affiliation{Department of Physics, Technical University of Munich, Garching D-85747, Germany}

\author{Jie Meng}
\email{mengj@pku.edu.cn}
\affiliation{State Key Laboratory of Nuclear Physics and Technology, School of Physics, Peking University, Beijing 100871, China}

\date{\today}




\maketitle


As a crucial ingredient of the nucleon-nucleon ($NN$) interaction, the tensor force has an important impact on the structural and dynamical properties of the nuclear many-body system, particularly the properties of exotic nuclei far away from the stability, which are crucial for understanding the nucleosynthesis in nuclear astrophysics. Many efforts have been devoted to studying the influence of the tensor force in the effective $NN$ interaction in the nuclear medium within the configuration interaction shell model~\cite{2020-Otsuka-RevModPhys.92.015002} and the density functional theories (DFTs)~\cite{2014-Sagawa-PPNP}. Due to the difference of modeling the effective $NN$ interaction and the difficulty of determining the tensor-force strengths in finite nuclei, the tensor-force effects in nuclear physics still need to be fully settled.

Starting from the realistic $NN$ interactions well calibrated in free space, the microscopic \textit{ab initio} calculations provide a more promising way to study the tensor-force effects since no free parameters are introduced throughout the solution of the many-body methods. Progress has been made on the role of the tensor force in the binding of light nuclei~\cite{1997-Pudliner-PhysRevC.56.1720} and the symmetry energy of nuclear matter~\cite{2011-Vidana-PhysRevC.84.062801}. Still, insufficient attention has been paid on the studies of tensor-force effects from microscopic \textit{ab initio} calculations.

The relativistic Brueckner-Hartree-Fock (RBHF) theory plays a key role in addressing the challenge of \textit{ab initio} methods rooted in a relativistic framework~\cite{2019-Shen-PPNP}. The strong correlation in the many-body wave function induced by the realistic $NN$ interaction $V$ is absorbed into the effective $G$ matrix, which is obtained as the solution of the Thompson equation in the medium~\cite{Brockmann1990_PRC42-1965}. In addition to the advantages by solving nuclear many-body systems within a relativistic framework, the main strength of the RBHF theory lies in its inherent ability to account for a class of three-body forces which are important for satisfactorily reproducing the empirical saturation properties of nuclear matter~\cite{Brockmann1990_PRC42-1965,2008-LiZH-PhysRevC.77.034316}. The RBHF theory has been successfully applied to describe the properties of nuclear matter and finite nuclei~\cite{WANG-SB2021_PRC103-054319,Shen2017PRC,2023-ZouWJ-arXiv}. 

In light of the remarkable reliability of the relativistic \textit{ab initio} theory, the many-body predictions from RBHF theory have been utilized as pseudo-data to constrain the parameters in both non-relativistic and relativistic DFTs.
By applying the RBHF theory in neutron drops, a systematic and specific pattern in the evolution of spin-orbit splittings is observed~\cite{2018-Shen-PLB778.344}, which is found to have strong tensor-force effects. 
This provides a novel scheme for determining the tensor-force strengths and forms an important guide for the microscopic derivations of nuclear DFTs~\cite{2019-Shen-PPNP}.

In this work, we aim to directly study the tensor-force effects in nuclear many-body systems from relativistic \textit{ab initio} theory. As a first step we focus on the infinite nuclear matter, which provides an ideal laboratory accessible to different methods that allows detailed benchmark. The RBHF theory in the full Dirac space is adopted, which considers simultaneously the positive- and negative-energy solutions of the Dirac equation, thus avoiding the uncertainties suffered from RBHF calculations without negative-energy states. Details can be found in Refs.~\cite{WANG-SB2021_PRC103-054319, 2022-Wang-SIBO-PhysRevC.106.L021305}. To identify the effects of tensor forces, we perform the RBHF calculations, in which the tensor forces in different meson-nucleon couplings are removed from the realistic $NN$ interaction by the non-relativistic reduced operators~\cite{2018-WangZH-PhysRevC.98.034313}
\begin{equation} \label{Vtphi}
  \hat{\mathcal{V}}^t_\phi = \frac{1}{m^2_\phi + \bm{q}^2} \mathcal{F}_{\phi} S_{12}.
\end{equation}
Here $\phi$ denotes the exchanged meson with mass $m_\phi$, $\bm{q}$ is the momentum transfer, and $S_{12}$ is the tensor force operator
\begin{equation}
  S_{12} \equiv (\bm{\sigma}_1\cdot\bm{q}) (\bm{\sigma}_2\cdot\bm{q}) - \frac{1}{3}(\bm{\sigma}_1\cdot\bm{\sigma}_2)q^2.
\end{equation}
The factor $\mathcal{F}_\phi$ in each meson-nucleon coupling can be found in Table 1 of Ref.~\cite{2018-WangZH-PhysRevC.98.034313}.
To study the tensor-force effects more precisely, a quenching factor $\lambda$ is introduced in Eq.~\eqref{Vtphi} for each coupling channel $\mathcal{F}_{\phi}$ to control the strength of the tensor force. The factor $\lambda=0$ means no tensor force, while $\lambda=1$ means the tensor forces are considered fully.

\begin{figure*}[t]
  \centering
  \includegraphics[height=5.5cm]{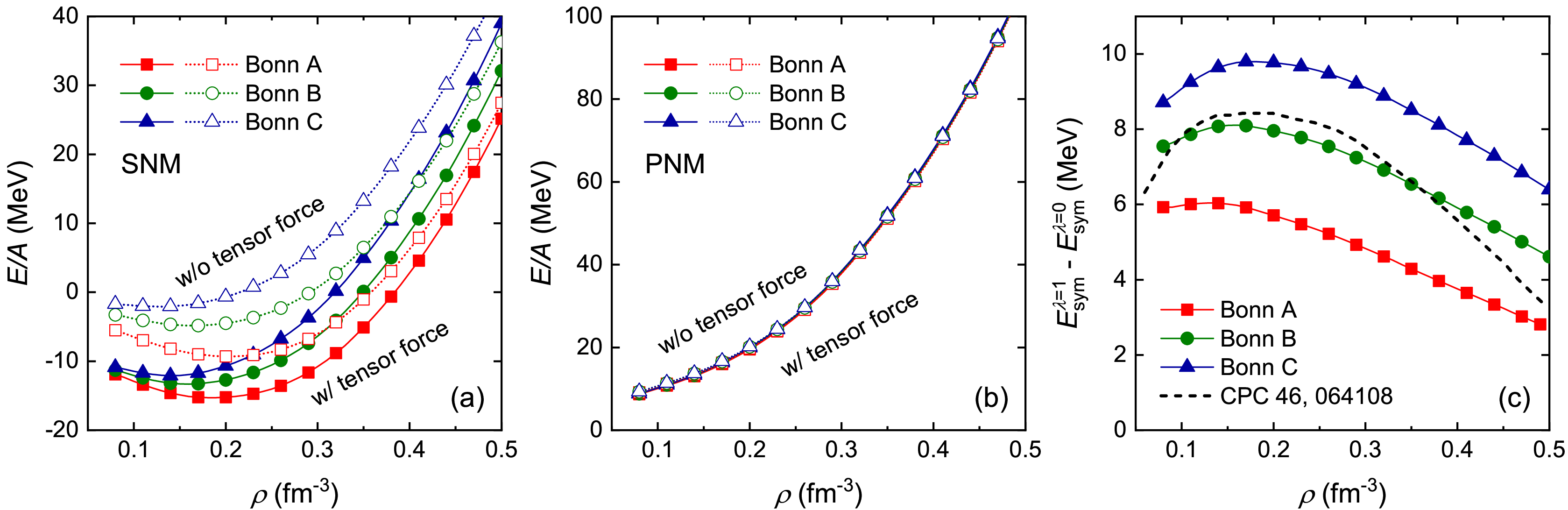}
  \caption{(Color online) Tensor-force effects on nuclear matter. The binding energies per particle of (a) SNM and (b) PNM calculated with (solid lines and full symbols) and without (dotted lines and empty symbols) tensor forces. (c) The symmetry energy difference obtained by performing the RBHF calculations with and without tensor force. The results from Ref.~\cite{2022-CCWang-Chin.Phys.C} are also shown for comparison. Details can be found in the text.}
  \label{label-fig1}
\end{figure*}

Figure~\ref{label-fig1}(a) shows the binding energies per particle as a function of the density $\rho$ for symmetric nuclear matter (SNM) calculated within the RBHF theory with and without tensor forces. The tensor-force effects lead to attraction over a wide range of densities and are more significant at lower densities. The characteristics of attraction can be understood from the Thompson equation, which can be written simply in the operator form $G = V + V (Q/e) V + \cdots$. Here $Q$ and $e$ denote the Pauli operator and the energy denominator, respectively. The tensor force in the first term $V$ does not contribute to the binding energy of nuclear matter due to the spin saturation. For the second term $V(Q/e)V$, due to the positive definiteness of quadratic terms of $V$ and the negative energy denominator in the presence of the Pauli operator, the contributions of the tensor force in the second term are always attractive. This implies that the contribution of tensor forces in realistic $NN$ interaction to the nuclear matter binding is reflected through higher-order effects.
In addition, among the three parameterizations of the Bonn potential~\cite{Brockmann1990_PRC42-1965}, the tensor-force effects of Bonn C are the most pronounced. This is consistent with the well-known fact that Bonn C has the strongest tensor force~\cite{Brockmann1990_PRC42-1965}, as indicated by the most significant $D$-state probability of the deuteron.

In Fig.~\ref{label-fig1}(b), we illustrate the tensor-force effects on the binding energies per particle in pure neutron matter (PNM) where the total isospin $T$ for two-nucleon state must equal $1$.
For the Bonn A potential, the calculations with and without tensor forces lead to nearly the same results.
This indicates that the total tensor-force effects for Bonn A in the isospin triplet with $T=1$ are marginal.
Similar conclusions can be drawn for Bonn B and Bonn C potentials.
This can be understood because the PNM is dominated by $^1S_0$ partial wave, while the tensor-force contribution appears in $P$ and higher waves.

The results in SNM and PNM allow us to derive the nuclear symmetry energy $E_{\text{sym}}(\rho)$, which is one of the cutting‐edge topics in nuclear physics and astrophysics. We depict the differences between the symmetry energies obtained with and without tensor forces in Fig.~\ref{label-fig1}(c).
The tensor force is essential around the empirical saturation density $\rho_0 = 0.16\pm0.01\ \text{fm}^{-3}$, beyond which the tensor-force effects weaken. 
The differences among Bonn potentials show the uncertainties from realistic $NN$ interactions.
Compared to our results, the symmetry energy difference obtained in Ref.~\cite{2022-CCWang-Chin.Phys.C} is very similar, where the tensor-force effects are analyzed by performing the RBHF calculations without the off-diagonal $NN$ matrix elements in the coupled spin-triplet channels, with the pvCD-Bonn potential~\cite{2019-WANGCC-Chin.Phys.C} as well as the projection techniques adopted.
By using the Hellmann-Feynman theorem and a non-relativistic \textit{ab initio} method with two- and three-body interactions~\cite{2011-Vidana-PhysRevC.84.062801}, the estimation of the tensor-force contribution to the symmetry energy is qualitatively consistent (36 MeV at $\rho=0.187\ \text{fm}^{-3}$), despite the quantitative discrepancy which may result from the $NN$ interactions, the many-body method, and the scheme to extract tensor-force effects.

It should be noted that starting from the effective $NN$ interaction within the DFTs, where only the tree-level interactions are considered, tensor force plays no role in nuclear matter due to the spin saturation. This reveals the advantages from \textit{ab initio} studies that can explore the higher-order effects of the tensor force on nuclear matter.


\begin{figure*}[t]
  \centering
  \includegraphics[height=5.5cm]{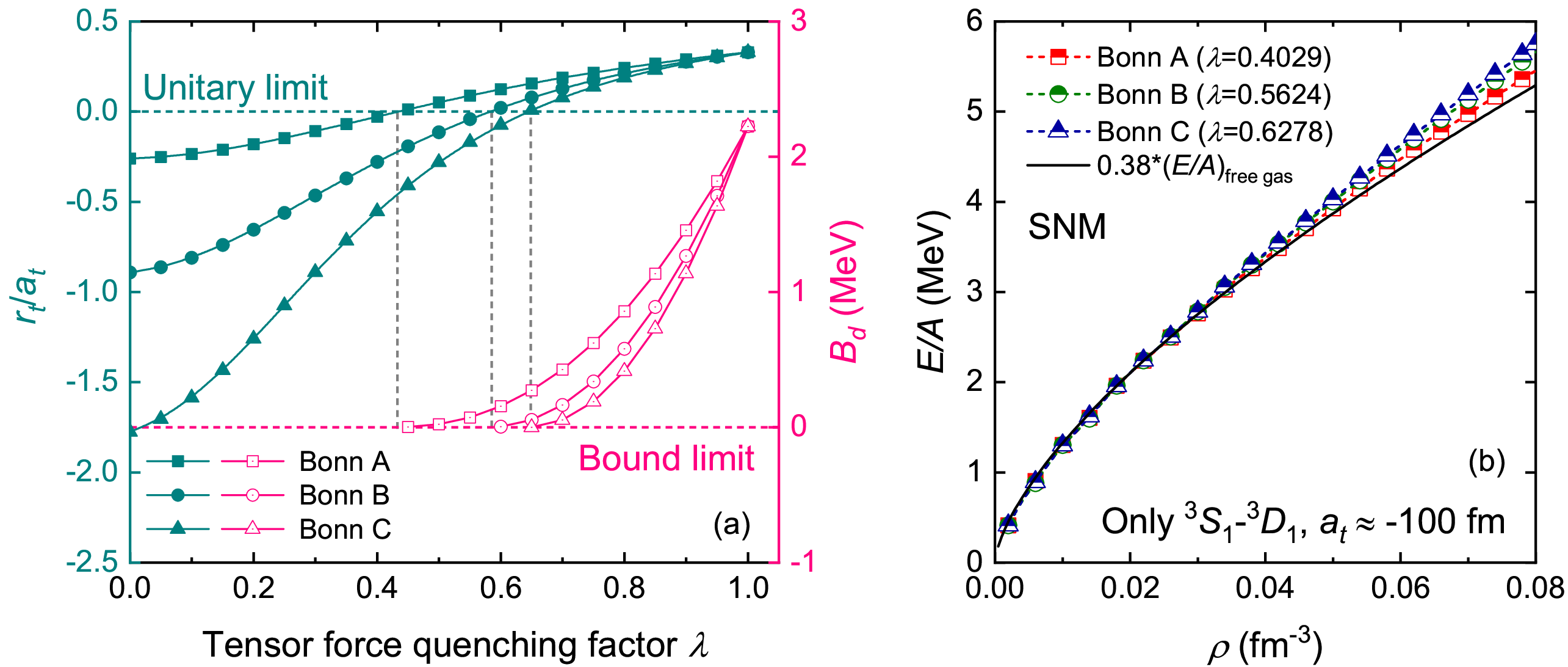}
  \caption{(Color online) (a) The dimensionless ratio $r_t/a_t$ in spin triplet (left axis) and the binding energy of the deuteron $B_d$ (right axis) with variation of the tensor force quenching factor $\lambda$. The upper (lower) horizontal dashed line denotes the locale of the unitary (bound) limit. (b) The ground-state energy per particle of SNM calculated with Bonn A, B, and C, where the scattering length $a_t$ in spin-triplet channel is tuned to $-100$ fm. Only contributions from the ${}^3S_1$-${}^3D_1$ channel are considered. The results of a free Fermi gas with a scaling factor 0.38 is also displayed for comparison.}
  \label{label-fig2}
\end{figure*}

The dilute fermion system at the unitary limit with infinite two-body scattering length is of general interest. The properties of the system exhibit universalities, regardless of the nature of constituent particles and their mutual interactions. In particular, the ground-state energy is proportional to that of a free Fermi gas, $E/A=\xi(E/A)_{\text{free\ gas}}$, where the scaling factor $\xi$ is called the Bertsch parameter~\cite{1999-Baker-PhysRevC.60.054311}. Such studies are of close relevance to various systems in condensed matter physics, nuclear physics, etc.

Nuclear physics is naturally in proximity to the unitary limit. This can be recognized because the scattering lengths in $^1S_0$ (spin singlet) and $^3S_1$ (spin triplet) are appreciably large compared to the effective range of the interaction. Many efforts have been devoted to studying the dilute PNM, which is dominated by the $^1S_0$ partial wave with the scattering length $a_s^{nn} = -18.9$ fm. By adjusting the mass of the sigma meson in the Bonn potentials, we tune the scattering length $a_s^{nn}$ to $-100$ fm and find the Bertsch parameter is 0.576. This value is close to the results from non-relativistic \textit{ab initio} calculations~\cite{2021-Vidana-10.3389/fphy.2021.660622}. The difference from the experimental measurements on ultra-cold atoms, 0.376(4)~\cite{2012-Ku-science.1214987}, might be related to the non-zero effective range of the neutron-neutron ($nn$) interaction~\cite{2011-Carlson-PhysRevA.84.061602}.

In comparison to the widely studied two-component (spin-$1/2$) unitary Fermi gas, here we generalize to consider a four-component (spin-$1/2$ and isospin-$1/2$) Fermi gas at zero temperature, since the isospin degree of freedom plays a vital role in nuclear physics. It is still not known whether the four-component unitary Fermi gas shares the same Bertsch parameter with the two-component unitary Fermi gas.

Different from the PNM, the SNM incorporates additionally the spin-triplet ${}^3S_1$ partial wave, where the scattering length $a_t=5.4$ fm implying the existence of a bound state for the neutron-proton $(np)$ system, i.e., the deuteron. It is well known that the tensor force plays a vital role in the deuteron's binding energy and leads to the coupling between ${}^3S_1$ and ${}^3D_1$ partial waves. In Fig.~\ref{label-fig2}(a) we present the ratio of the effective range $r_t$ and the scattering length $a_t$ as a function of the strength of the tensor force. With the continuously decreasing of the quenching factor $\lambda$ of the tensor force, the ratio $r_t/a_t$ also decreases. Note that the running of $r_t/a_t$ is dominated by the scattering length $a_t$, while $r_t$ retains in the range of 1.8-5.1 fm. At a certain $\lambda$, the scattering length $a_t$ goes to infinity, locating the $np$ system at the unitary limit, as denoted by the upper horizontal dashed line. This also illustrates that the strong tensor force in the $np$ interaction makes the $np$ system somewhat deviate from the unitary limit.

We also show the tensor-force effects on deuteron's binding energy in Fig.~\ref{label-fig2}(a). With the decrease of the factor $\lambda$, the binding energy of the deuteron $B_d$ also decreases from 2.2246 MeV to zero until meeting the unitary limit. The most rapid case is found for Bonn C, since this potential has the strongest tensor force. The deuteron is not bound any more after the dimensionless ratio $r_t/a_t$ crosses the locale of the unitary limit and the bound limit is denoted by the lower horizontal dashed line.

It is interesting to study how the dilute SNM with only the coupled spin-triplet ${}^3S_1$-${}^3D_1$ partial waves behave at the unitary limit.
By tuning the quenching factor of the tensor force $\lambda=0.4029, 0.5624$, and $0.6278$ for Bonn A, B, C, respectively, we obtain the scattering length $a_t\approx -100$ fm, whose magnitude is large enough compared to the effective range $r_t=2.5$\ fm. From the modified Bonn potentials, we calculate the ground-state energies per particle of SNM by only considering the $NN$ matrix elements in the ${}^3S_1$-${}^3D_1$ channel. The Thompson equations are solved only once with initial vanishing single-particle potentials to avoid the numerical instabilities at extreme low densities. This should be reasonable since the hypothetically converged single-particle potentials are very close to zero at extremely low densities.

In Fig.~\ref{label-fig2}(b), we present the ground-state energy per particle of SNM within the scheme discussed above.
The results from Bonn A, B, and C show very good consistency, not sensitive to the details of the interaction.
Furthermore, it is found that the results from our calculations are proportional to that of a free Fermi gas for lower densities. The aforementioned Bertsch parameter is determined as $\xi=0.38$.
We also obtain the Bertsch parameter $\xi' = 0.376$ for dilute SNM in the $^1S_0$ channel, where the scattering length of the $np$\ ($nn$, proton-proton) system is tuned to $-100$ fm by adjusting the mass of the sigma meson. The almost same Bertsch parameter for dilute SNM in $^1S_0$ and ${}^3S_1$-${}^3D_1$ channels may be related to Wigner's SU(4) symmetry~\cite{1937-Wigner-PhysRev.51.106}, since the scatting lengths in the $^1S_0$ and the ${}^3S_1$-${}^3D_1$ are the same ($-100$ fm) after the adjustment. 

Particularly, in the $^1S_0$ channel, the Bertsch parameter 0.376 for dilute SNM is smaller than the value 0.576 for dilute PNM, due to the additional $np$ interaction.
This indicates that the Bertsch parameter of four-component unitary Fermi gas is different from the two-component case.
Our calculations for the four-component unitary Fermi gas with spin and isospin can provide valuable reference to other theoretical studies and ultra-cold atom experiments on four-component fermion systems.

In summary, the tensor-force effects on the equation of state of nuclear matter and the symmetry energy have been studied from the realistic Bonn potentials, by invoking the RBHF theory with the tensor-force matrix elements subtracted.
For the binding energies per particle of SNM and the symmetry energy, the tensor-force effects are attractive and are more pronounced around the empirical saturation density. For PNM, the tensor-force effects are marginal.
The impact of the tensor force on the deviation of the $np$ system to the unitary limit has also been studied. An infinite (negative) scattering length in the spin triplet can be obtained by tuning the tensor-force strength, and the dilute SNM is located at the unitary limit.
With only the interaction in the ${}^3S_1$-${}^3D_1$ channel considered, the ground-state energy of dilute SNM is found proportional to that of a free Fermi gas with a scaling factor 0.38, which reveals very good universal properties.
By applying the method to analyze the tensor-force effects, this work paves the way to study the tensor-force effects in neutron stars as well as finite nuclei from realistic $NN$ interactions. This work also highlights the role of the tensor force on the deviation of the nuclear physics to the unitary limit and provides valuable reference for studies of the four-component unitary Fermi gas.

{\bf Conflict of interest}

The authors declare that they have no conflict of interest.

{\bf  Acknowledgments}

S. Wang and Q. Zhao thank S. Aoki for helpful discussions. This work was supported in part by the National Natural Science Foundation of China (NSFC) under Grants No. 12205030, No. 12347101, No. 11935003, No. 11975031, No. 11875075, No. 12070131001 and No. 12047564, the National Key R\&D Program of China under Contracts No. 2017YFE0116700 and No. 2018YFA0404400, the Fundamental Research Funds for the Central Universities under Grants No. 2020CDJQY-Z003 and No. 2021CDJZYJH-003, the MOST-RIKEN Joint Project "Ab initio investigation in nuclear physics", the Institute for Basic Science under Grant No. IBS-R031-D1, and the Deutsche Forschungsgemeinschaft (DFG, German Research Foundation) under Germany’s Excellence Strategy EXC-2094-390783311, ORIGINS. 
Part of this work was achieved by using the supercomputer OCTOPUS at the Cybermedia Center, Osaka University under the support of Research Center for Nuclear Physics of Osaka University and the High Performance Computing Resources in the Research Solution Center, Institute for Basic Science.

{\bf Author contributions}

Sibo Wang and Qiang Zhao conceived the idea. Sibo Wang performed numerical calculations and wrote the original draft. All authors contributed to discussions of results and finalized manuscript.


\end{document}